\documentclass[aps, prl, preprint, superscriptaddress]{revtex4-1} 
\usepackage{graphicx}
\usepackage{amsmath}
\usepackage{amssymb}

\usepackage{times}

\usepackage[english]{babel}
\usepackage{siunitx}
\usepackage{color}

\newcommand{\abs}[1]{\left|#1\right|}
\newcommand{\ket}[1]{\left|#1\right>}

\newcommand{\ba}{\textbf{a}}
\newcommand{\bH}{\textbf{H}}
\newcommand{\bL}{\textbf{L}}
\newcommand{\bD}{\textbf{D}}

\newcommand{\NN}{\mathcal N}

\begin{document} 

\title{Coherent oscillations inside a quantum manifold stabilized by dissipation}

\author{S. Touzard}
\email{steven.touzard@yale.edu}
\affiliation{Department of Applied Physics and Physics, Yale University, New Haven, CT 06520, USA}
\author{A. Grimm}
\affiliation{Department of Applied Physics and Physics, Yale University, New Haven, CT 06520, USA}
\author{Z. Leghtas}
\affiliation{Department of Applied Physics and Physics, Yale University, New Haven, CT 06520, USA}
\author{S.O. Mundhada}
\affiliation{Department of Applied Physics and Physics, Yale University, New Haven, CT 06520, USA}
\author{P. Reinhold}
\affiliation{Department of Applied Physics and Physics, Yale University, New Haven, CT 06520, USA}
\author{C. Axline}
\affiliation{Department of Applied Physics and Physics, Yale University, New Haven, CT 06520, USA}
\author{M. Reagor}
\affiliation{Department of Applied Physics and Physics, Yale University, New Haven, CT 06520, USA}
\author{K. Chou}
\affiliation{Department of Applied Physics and Physics, Yale University, New Haven, CT 06520, USA}
\author{J. Blumoff}
\affiliation{Department of Applied Physics and Physics, Yale University, New Haven, CT 06520, USA}
\author{K.M. Sliwa}
\affiliation{Department of Applied Physics and Physics, Yale University, New Haven, CT 06520, USA}
\author{S. Shankar}
\affiliation{Department of Applied Physics and Physics, Yale University, New Haven, CT 06520, USA}
\author{L. Frunzio}
\affiliation{Department of Applied Physics and Physics, Yale University, New Haven, CT 06520, USA}
\author{R.J. Schoelkopf}
\affiliation{Department of Applied Physics and Physics, Yale University, New Haven, CT 06520, USA}
\author{M. Mirrahimi}
\affiliation{Yale Quantum Institute, Yale University, New Haven, Connecticut 06520, USA}
\affiliation{QUANTIC team, INRIA de Paris, 2 Rue Simone Iff, 75012 Paris, France}
\author{M.H. Devoret}
\affiliation{Department of Applied Physics and Physics, Yale University, New Haven, CT 06520, USA}

% Include the date command, but leave its argument blank.

\date{\today}


\begin{abstract}
Manipulating the state of a logical quantum bit usually comes at the expense of exposing it to decoherence.
Fault-tolerant quantum computing tackles this problem by manipulating quantum information within a stable manifold of a larger Hilbert space, whose symmetries restrict the number of independent errors. The remaining errors do not affect the quantum computation and are correctable after the fact. Here we implement the autonomous stabilization of an encoding manifold spanned by Schr\"odinger cat states in a superconducting cavity. We show Zeno-driven coherent oscillations between these states analogous to the Rabi rotation of a qubit protected against phase-flips. Such gates are compatible with quantum error correction and hence are crucial for fault-tolerant logical qubits. 
\end{abstract}

\maketitle

\begin{figure*}
	\includegraphics[height=13cm]{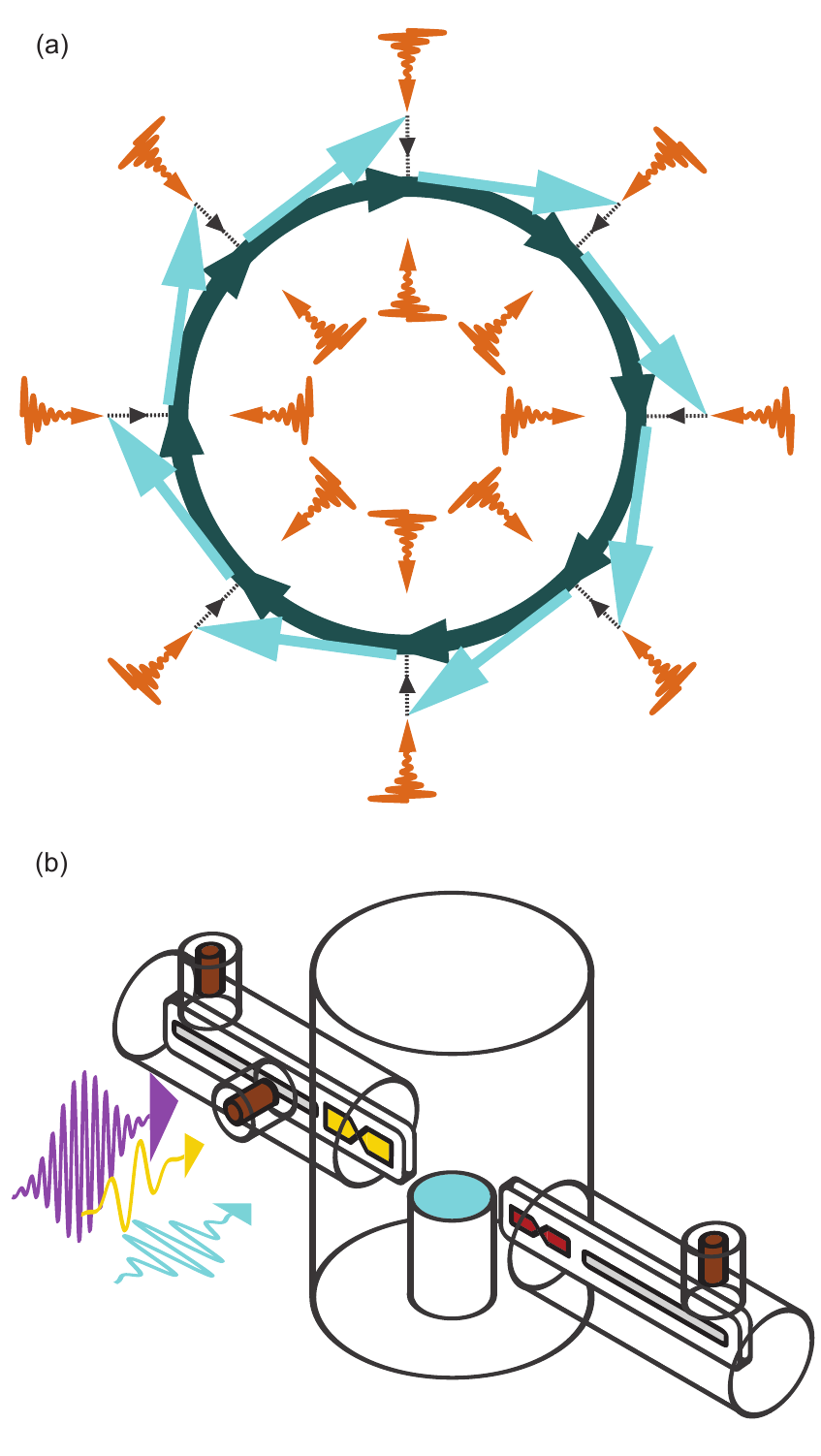}
	\caption{
		Quantum Zeno dynamics and its implementation. (a) Conceptual representation of the experiment. The quantum state of a harmonic oscillator is represented here by a point in a 2D plane (not to be confused with phase space). The dark blue circle represents a cross-section  of the Bloch sphere of a two-state manifold in the larger Hilbert space of the oscillator. The quantum Zeno effect observed in our experiment corresponds to motion along the circle. A weak excitation drive is applied to the oscillator and the resulting trajectory has both a component along the circle and out of it. The non-linear dissipation and drive (orange) cancels the movement outside the circle while being blind to the position of the quantum state on the circle.
		(b) Schematics of the experimental device. The quantum manifold is stabilized within the Hilbert space of the fundamental mode of an aluminium post cavity (cyan, storage in the text). This resonator is coupled to two Josephson junctions on sapphire (yellow for the reservoir and crimson for the transmon qubit, see text), which are read out by stripline resonators (grey). Three couplers (brown) bring microwave drives into the system and carry signals out of it.
	}
	\label{fig:Fig1}
\end{figure*}

%Intro on QZE
The quantum Zeno effect (QZE) is the apparent freezing of a quantum system in one state under the influence of a continuous observation. This continuous observation can be performed by a dissipative environment \cite{Sudarshan1977, Wineland1990, Raizen1997}. It can be further generalized to the stabilization of a manifold spanned by multiple quantum states, an operation which requires a dissipation that is blind to the manifold observables \cite{Sudarshan2000}. Harnessing this effect is crucial for the design of quantum computation schemes, since autonomous stabilization is a form of the feedback needed for quantum error correction. When employing manifold QZE for correcting errors, motion inside the manifold can still subsist and can be driven by the combination of the dissipative stabilization and an external force \cite{Knight2000, Vedral2006, Calsamiglia2010, ZanardiCampos2014, Jiang2016, Rouchon2017}. Therefore manifold QZE offers a pathway towards the realization of logical gates compatible with quantum error correction.
%Introducing the stabilization and the oscillations
%A dissipation mediated by pairwise exchange of excitations between oscillator and a cold bath realizes this goal. 
An example of such a system is provided by a superconducting microwave cavity, in which a dissipative process that annihilates photons in pairs at rate $\kappa_2$, acting together with a two-photon drive of strength $\epsilon_2$ projects the system onto the manifold spanned by Schr\"odinger cat states $\ket{\mathcal{C}_{\alpha_\infty}^{\pm}}=\NN\left(\ket{{\alpha_\infty}}\pm\ket{-{\alpha_\infty}}\right)$, where $\ket{{\alpha_\infty}}$ is a coherent state of amplitude $\alpha_\infty = \sqrt{2\epsilon_2/\kappa_2}$ and $\NN$ is a normalization factor  \cite{Carmichael1988, Devoret2014, Devoret2015}. Each one of these states has a well defined photon number parity, which is conserved by the engineered dissipation. In this Schr\"odinger cat states manifold, the displacement operator $\bD(\alpha) = \exp(\alpha\ba^\dag - \alpha^*\ba)$ (where $\ba$ is the annihilation operator acting on the harmonic oscillator) has two effects: it changes the photon number parity and changes the amplitude of its component coherent states. The engineered dissipation leaves the change in parity invariant and cancels the change in amplitude (Fig~\ref{fig:Fig1}a). The net result of this Quantum Zeno Dynamics (QZD) is to continuously vary the parity of Schr\"odinger cat states.

% Quantum information POV
These parity oscillations constitute the basis of an X-gate on a qubit encoded in the protected manifold $\ket{0/1}_P = \NN\left(\ket{{\alpha_\infty}}\pm\ket{-{\alpha_\infty}}\right)$. Encoding quantum information in superpositions of Schr\"odinger cat states is compatible with quantum error correction (QEC) realized with quantum non-demolition parity measurements \cite{Davidovich1997, Schoelkopf2014, Schoelkopf2016_2}. Our gate is fundamentally different than previous manipulations of Schr\"odinger cat states \cite{Schoelkopf2016_3} as it operates while the manifold is stabilized. Thus, the quantum information is protected from out-of-manifold gate errors. Moreover the operation of the gate is not affected by the dominant source of errors: bit-flips. In fact, as the operation commutes with them, it is compatible with a fault-tolerant scheme that would correct them after the operation.

While related driven manifold dynamics have been proposed and observed \cite{Pascazio2002, Haroche2012, Smerzi2014, Brune2014, Huard2015}, the non-linear dissipation specific to our experiment adds a crucial element: any drift out of the cat state manifold is projected back into it.

% Second paragraph 
% Introduction to the 2 mode hamiltonian and phase:
% - The dynamics is made by two photon exchanges between S and R
% - Those exchanges are stimulated by a microwave pump at frequency BLA and at frequency omega_R
% - The phase of beta is set by the phase of the stabilization drives
% - We lock them to the phase of the storage to make it real
% - The linear drive is most effective when its phase is purely imaginary

In our experiment, schematically shown in Fig~\ref{fig:Fig1}, the drive-dissipation is implemented using two-photon transitions between two electromagnetic modes. The first one has a high quality factor and stores the Schr\"odinger cat states. We refer to it as the storage (subscript S). The second one is used as an engineered cold bath that removes rapidly the entropy from the storage. We refer to it as the reservoir (subscript R). We employ the four-wave mixing capability of a Josephson junction, together with two microwave pumps, to stimulate those transitions. In order to make resonant the conversion from one reservoir photon into two storage photons and vice-versa, the first pump is set at frequency $2f_S - f_R$. The second pump, set at frequency $f_R$, combines with the first one to create pairs of photons in the storage. When the dynamics of the reservoir mode is eliminated, the density matrix of the storage mode $\rho$ is given by the Lindblad equation 
\begin{equation*}
\frac{d\rho}{dt} = -\frac{i}{\hbar}[\bH_S, \rho] + \frac{\kappa_2}{2} \mathcal{D}[\ba_S^2-{\alpha_\infty}^2]\rho,
\end{equation*}
where $\bH_S$ is a Hamiltonian acting on the storage and $\mathcal{D}[\bL]\rho = 2\bL\rho\bL^\dag-\bL^\dag\bL\rho - \rho\bL^\dag\bL$ is the Lindblad superoperator. As the Lindblad superoperator is engineered to be the dominant term in the dynamics, the dynamical steady states of the system are given by the coherent states $\ket{\pm {\alpha_\infty}}$. The microwave pumps set the phase and amplitude of the complex amplitude $\alpha_\infty$. The Hamiltonian part of the equation contains the self Kerr effect of the storage mode induced by the Josephson junction and the linear drive that induces the coherent oscillations: $\bH_S/\hbar = -\chi_{SS}/2 (\ba_S^{\dag 2}\ba_S^2) + (\epsilon \ba_S^\dag + \rm{h.c.})$. The frequency of the coherent oscillations is maximum when the phase of the linear drive $\epsilon$ is perpendicular to the phase of the stabilized Schr\"odinger cat states. 
% Set orders of magnitude inside this equation
Thus, this linear drive displaces the Schr\"odinger cat state perpendicularly to the stabilization axis while the dissipation continuously projects the system back to the stabilized manifold. If the drive respects the adiabaticity condition $\abs{\epsilon} \ll \abs{{\alpha_\infty}}^2\kappa_2$ then the net effect of the linear drive is to induce parity oscillations within the stabilized manifold, at frequency $\Omega = 2\epsilon\abs{{\alpha_\infty}}$ \cite{Devoret2014}.

% Architecture and its components

% Third paragraph (Storage)

% Now that we have all the terms we can compare them together and give numerical values
% - Storage is a microwave post cavity
% - Single photon dissipation rate and Kerr (killed by kappa 2)
% - Adiabaticity condition 

% How the architeture helped the coherence
The adiabaticity condition \cite{Vedral2006, Calsamiglia2010, ZanardiCampos2014, Jiang2016, Rouchon2017} sets an upper-bound on the frequency of these oscillations, fixed by the maximum $\kappa_2$ that we can engineer. In order to observe this dynamics, we also need the coherence time of the storage mode to be larger than the period of the oscillations. The architecture we designed was key to engineer both a highly coherent storage mode and a large coupling to the environment. 
% Explain the storage that is highly coherent
We implement the storage mode into a long-lived post cavity made of aluminium  \cite{Schoelkopf2013_2}(Fig~\ref{fig:Fig1}b). Its finite lifetime induces two types of errors on a protected qubit encoded in the stabilized manifold. First, in absence of stabilization, the amplitude of the Schr\"odinger cat  states decays until eventually the two coherent states are no longer distinguishable. This error happens at rate $\kappa_1/2\pi =$\SI{1.7}{\kilo \hertz}. Second, when the environment is observed to have absorbed a photon, the projected density matrix of the storage mode suffers a parity jump, which corresponds to a bit-flip error in our encoded qubit. For a stabilized cat state containing $\bar{n}=\abs{{\alpha_\infty}}^2$ photons on average, they happen at a rate $\bar{n}\kappa_1$ \cite{Haroche2006}. Additionally, the above mentioned Kerr effect would distort the coherent states at rate $\chi_{SS}/2\pi\sim$\SI{3}{\kilo \hertz} in absence of stabilization. 
% Now the coupling to the environment is high
In order to achieve a fast non-linear dissipation, the storage cavity is coupled to a transmon qubit embedded into a coaxial tunnel \cite{Schoelkopf2016_4}, whose lifetime is engineered to be much less than that of the storage (\SI{317}{\nano \second}), and we use it as the entropy reservoir to induce the QZE. While the reservoir is efficient to dispose of the entropy of the storage mode, its low lifetime has two impacts on the coherence of the storage. First, we associate the lifetime of our storage cavity  to the Purcell effect (usually $\kappa_1/2\pi$ is of order \SI{100}{\hertz} \cite{Schoelkopf2013_2}). Second the finite temperature of the reservoir causes additionnal dephasing of the storage mode (see supplement \cite{supp}). However, the direct coupling between the storage and the reservoir modes led to a non-linear dissipation rate of $\kappa_2/2\pi = $\SI{176}{\kilo \hertz}. The two orders of magnitude separating $\kappa_1$ and $\kappa_2$ are enough to observe the parity oscillations while respecting the adiabaticity condition.
% Very coherent transmon for measurement
The storage cavity is also coupled to a very coherent transmon whose coherence times ($T_1=$ \SI{70}{\micro \second}, $T_2^*=$ \SI{30}{\micro \second}) are large compared to the time it takes to perform a parity measurement of the storage cavity using the dispersive coupling (\SI{218}{\nano \second}). We use this transmon qubit to measure the Wigner function of the storage mode.

\begin{figure*}
	\includegraphics[height=13cm]{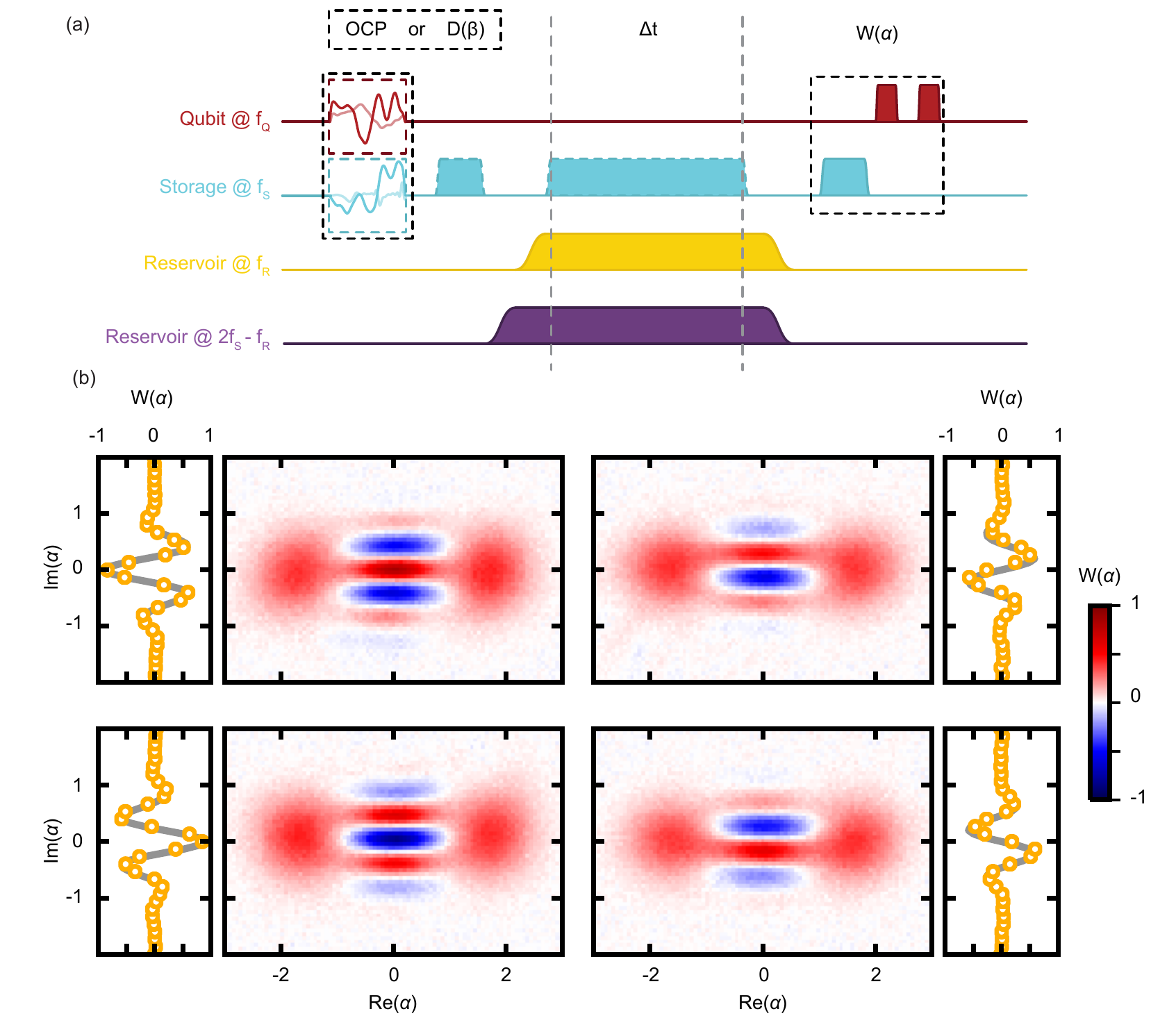}
	\caption{
		Experimental protocol and Wigner tomography result. (a) Sequence of different drives outlined in fig 1. The storage is either initialized in a cat state $\NN\left(\ket{{\alpha_\infty}}\pm\ket{-{\alpha_\infty}}\right)$ with optimal control pulses or in a coherent state $\ket{\pm{\alpha_\infty}}$ with a displacement D($\pm {\alpha_\infty}$). The drives stabilizing the manifold are turned on in \SI{24}{\nano \second} (purple and yellow). They are on for a duration $\Delta t$ during which the storage drive (cyan) can be turned on to induce the parity oscillations. The drives are left on for another \SI{500}{\nano \second} and then turned off in \SI{24}{\nano \second} after which the Wigner function is measured. (b) In the left column is the Wigner functions of the storage cavity after initialization in an even or odd cat state $(\abs{{\alpha_\infty}}^2=3)$. The right column shows the corresponding Wigner functions after a quarter of an oscillation. The colormaps are averaged raw data of the Wigner function measurement (see text) and the orange circles are cuts along Re($\alpha$)=0. The grey solid lines are theoretical curves corresponding to even or odd cats (left column, lines 1 and 2 respectively) and parityless cats (right column, lines 1 and 2). The only fit parameter in the theory is the renormalization of the amplitude by a factor 0.87 on the left, and 0.65 on the right. These factors account for the fidelity of the parity measurement and the decay of the fringes of the cat states during the stabilization. 
	}
	\label{fig:Fig2}
\end{figure*}

Our experimental protocol follows a fixed sequence of pulses which contains three parts (Fig~\ref{fig:Fig2}a). The first step is the initialization of the system in the encoding manifold, which is done with pulses generated by an optimal control algorithm \cite{Schoelkopf2016_3}. As it involves transient states that are not Schr\"odinger cat states and that are entangled to the transmon qubit, this method induces errors on the protected encoding that are not corrected by the stabilization. However, it is currently the fastest method available. The second part is the stabilization of the manifold, which is done with or without the rotation drive. Finally, in the third part, the Wigner function of the storage cavity at a given point in phase space is measured \cite{Davidovich1997, Haroche2002}.

We characterized the initialization and the quality of the measurement by taking a full Wigner tomography of the storage cavity initialized in $\ket{0}_P$ and $\ket{1}_P$ (Fig~\ref{fig:Fig2}b). The raw data consisted of single shot parity measurements realized with a parametric amplifier and averaged without any further normalization. The phase locking of the different drives ensured that the stabilization axis of the Schr\"odinger cat states was aligned with the Wigner representation axis, and that the rotation drive was perpendicular to the stabilization axis \cite{supp}. The right column illustrates our ability to go from an even/odd parity Schr\"odinger cat state to a "Yurke-Stoler" cat state \cite{Yurke1985}. The zero value of the Wigner function at the center of phase space shows that these states had no parity. They were generated by a rotation of $\pi/2$ in the encoding manifold. It is important to note that the cat is not pushed sideway. The fringes are moving, but the "blobs" remain in place (Fig~\ref{fig:Fig2}b).

\begin{figure*}
	\includegraphics[height=11cm]{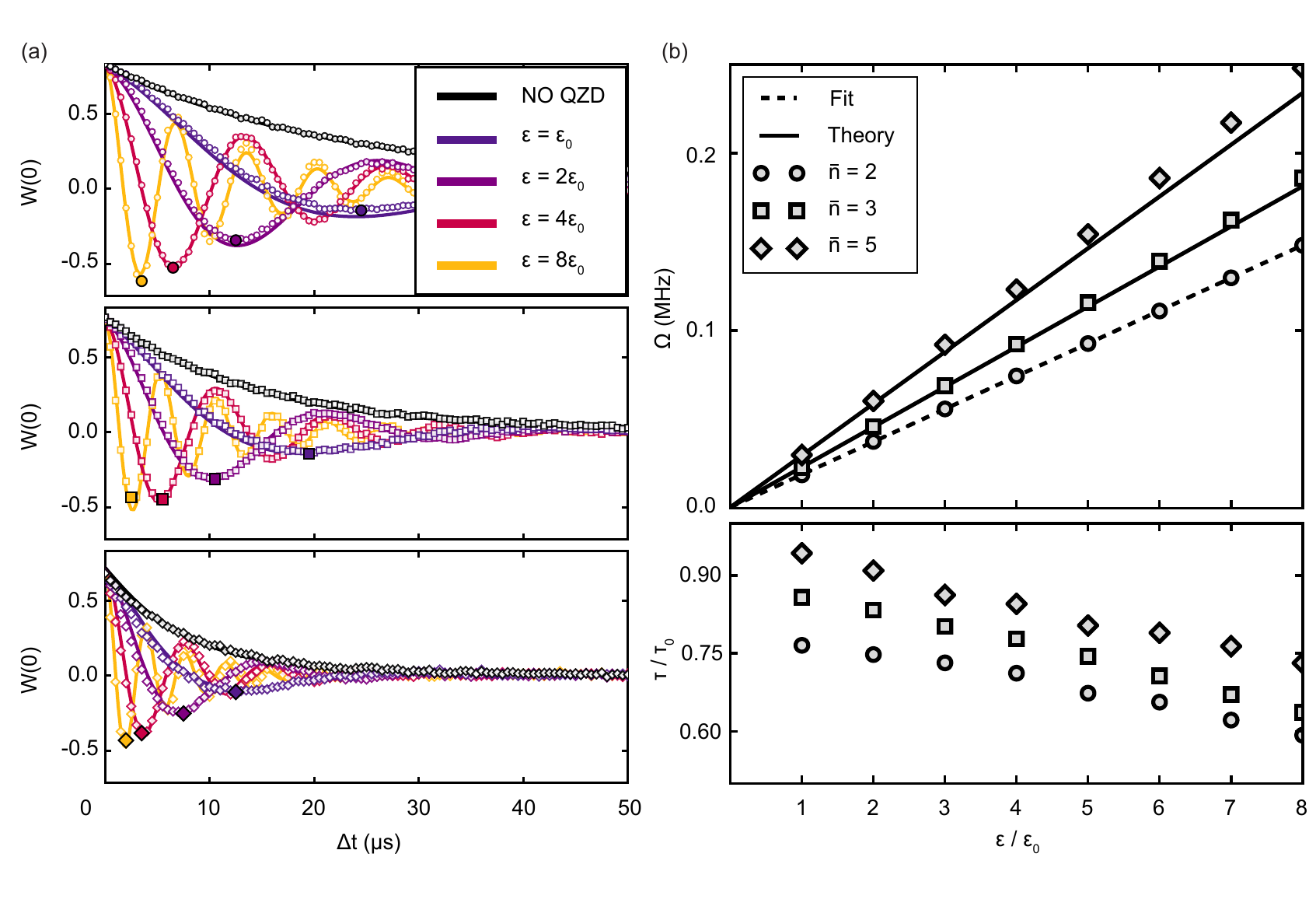}
	\caption{
		Characterization of the oscillations. (a) Evolution of the measured parity as a function of time. The initial cat states are even, with $\bar{n}=\abs{{\alpha_\infty}}^2=2, 3, 5$ (circles, squares, diamonds). The storage drive is either off (black markers) or on (coloured markers) with various strengths given in units of a chosen base strength $\epsilon_{0}$. Simulations are shown as solid lines. The minimum of each experimental curve is emphasized for each drive strength (full marker with black contour). (b) A fit of the data gives the frequency $\Omega$ and the time constant $\tau$ of the decaying oscillations. The former is plotted as a function of the relative drive strength $\epsilon / \epsilon_0$ (top panel). The case $\bar{n}$=2 is fitted with a linear function (dashed line). Based on this, we make predictions for $\bar{n}$=3, 5 (solid lines). The bottom panel shows the characteristic decay time of the oscillations $\tau$, normalized by the decay time of the non-driven case $\tau_0$}
	
	\label{fig:Fig3}
\end{figure*}

In order to investigate the parity oscillations more closely, we restricted the measurement of the Wigner function to the center of phase space (photon number parity measurement). In Fig~\ref{fig:Fig3}a we present the time evolution of cat states, initially in the even state. We measured their parity over \SI{50}{\micro \second} while they were stabilized (Fig~\ref{fig:Fig3}a). For $\bar{n} =$ 2, 3 and  5 we observed decay time constants of respectively \SI{22}{\micro \second}, \SI{14}{\micro \second} and \SI{8}{\micro \second}. 
This behaviour arises from the natural single-photon jumps of the cavity. They correspond to bit-flips within the encoding manifold which eventually destroy the coherence of the encoded qubit. The coherence of the encoded qubit is lost at a rate $2\bar{n}\kappa_1$ \cite{Haroche2006}. This is close to what was found in the experiment and thus shows that the decoherence is mainly due to bit-flips happening during the stabilization. With the rotation drive turned on, the oscillations of the parity over time are similar to Rabi oscillations for a two-level system. For a drive with strength $\epsilon$ the equivalent Rabi frequency  \cite{Devoret2014} is given by $\Omega = 2\epsilon\abs{{\alpha_\infty}}$. We chose a first drive strength $\epsilon_0$ that gave a single oscillation in parity within the decay time of a Schr\"odinger cat state with amplitude $\bar{n}=$ 2. We then repeated the experiment for drive strengths that were multiples of $\epsilon_0$ and for different amplitudes of the initial cat states. On each panel the frequency of the oscillations increases with the drive strength. By looking at curves that correspond to the same drive strength over different panels (same colour) we see that the frequency of oscillation also increases with the amplitude of the initial state. We obtained theory predictions by numerically integrating the evolution of the density matrix and superimposed them on the data. The parameters of the theory were all provided by the results of independent experiments  \cite{supp}.

We also present in Fig~\ref{fig:Fig3}b the frequency of the oscillations as a function of the normalized drive strength (Fig~\ref{fig:Fig3}b). According to theory, the oscillation frequency should depend linearly on the drive strength. The linear fit for $\bar{n} = 2$ gives $\epsilon_0/2\pi = $\SI{7}{\kilo \hertz}. This value, when compared to $\bar{n}\kappa_2$, means that we respected the adiabaticity condition for this drive strength. However, when the drive strength increases, this condition is no longer fulfilled.  Subsequently, we predict the oscillation frequencies for $\bar{n} = $ 3 and 5 with good agreement. The difference between the prediction and the data indicates that the stabilized cat state might have had a larger amplitude than the one measured. This is corroborated by the fact that the decoherence timescales for $\bar{n} =$ 3 and 5 were lower than predicted.

The second panel of Fig~\ref{fig:Fig3}b shows the evolution of the normalized decay time constant for different cat state amplitudes as a function of the drive strength. If the gate was infinitely slow, the decay constant would be the same as in the non-driven case. However, when $\epsilon/\epsilon_0$ increases, the oscillations decay faster. This is explained by the fact that the gate does not perfectly respect the adiabaticity condition  $\abs{\epsilon} \ll \abs{{\alpha_\infty}}^2\kappa_2$. Nevertheless, when the number of photons in the initial state is larger, the decay constant of the oscillations gets closer to the ideal limit: the adiabaticity condition is easier to fulfill for higher number of photons. Although encoding with a Schr\"odinger cat state of larger amplitude increases the  bit-flip rate, given by $2\bar{n}\kappa_1$, it increases the quality of our manipulation.

\begin{figure*}
	\includegraphics{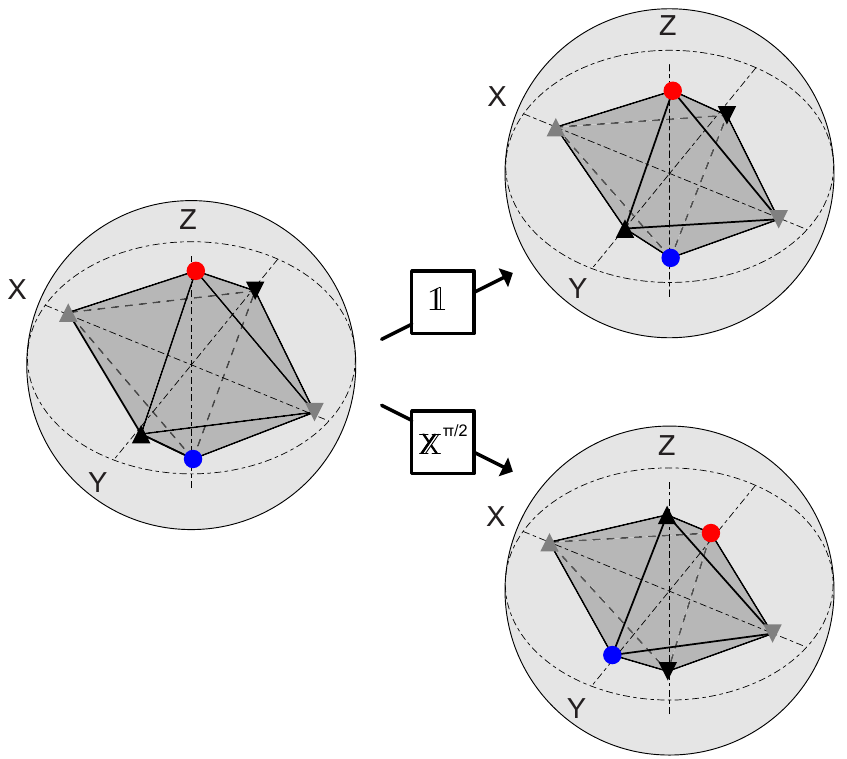}
	\caption{
		Gate on cardinal points of the Bloch sphere of the protected manifold. An arbitrary cat state $\NN\left(\cos(\frac{\theta}{2})\ket{\mathcal{C}_{{\alpha_\infty}}^{+}}+\sin(\frac{\theta}{2})e^{i\phi}\ket{\mathcal{C}_{{\alpha_\infty}}^{-}}\right)$ is represented by a point on a sphere. Six initial states are chosen, corresponding to the cardinal points $(\theta, \phi) = (0, 0), (\pi, 0), (\pm\pi/2, 0), (\pi/2, \pm\pi/2)$, with $\abs{{\alpha_\infty}}^2=3$, and their equivalent Pauli operators are measured. The markers corresponding to each initial state are respectively red and blue circles, grey up/down triangles, and black up/down triangles. The initial octahedron formed by those points (left) is either transformed under the action of the identity (upper right) or a rotation of $\pi/2$ around the X axis (lower right).
	}
	\label{fig:Fig4}
\end{figure*}

Finally, we measured the effect of this protected Rabi rotation on an arbitrary state of the encoding manifold. We represent a state of the protected qubit by a vector in the Bloch sphere. Its coordinates were found by measuring the equivalent Pauli operators for this encoding \cite{Schoelkopf2015}. The effect of the gate is accurately described by its effect on the 6 cardinal points of an octahedron within the Bloch sphere. We present the results for a manifold encoding using $\abs{{\alpha_\infty}}^2 = 3$ (Fig~\ref{fig:Fig4}). The octahedron on the left shows the initial state. To illustrate the effect of the gate we chose a specific rotation of $\pi / 2$ using a drive strength $\epsilon = 6 \epsilon_0$. This corresponded to a gate time of \SI{1.8}{\micro \second}. We compared the results with those obtained by waiting for the same amount of time without applying a drive, which corresponded to applying the identity.

The next step after manipulating an encoded and protected qubit contained in the stabilized manifold of cat states is to address the fault-tolerance of logical operations. A future version of our experiment, which should be accessible with current techniques, will increase $\kappa_2$ above any coupling to other modes and thus achieve two goals: first, it will improve the gate quality to make it better than a gate on a physical qubit. Second, it will suppress the dephasing due to finite temperature in other modes and thus suppress one remaining decoherence channel. All possible remaining errors would then be equivalent to bit-flip errors, which can be corrected by fault-tolerant joint parity measurements \cite{Mirrahimi2016, Schoelkopf2016} on several cavities.

\begin{acknowledgments}
We acknowledge Victor Albert and Liang Jiang for helpful discussions, Kyle Serniak and Luke Burkhart for their work on the fabrication process, R\'emi Bisognin and Renan Goupil for their participation in the experiment. Facilities use was supported by the Yale SEAS clean room, YINQE, and NSF MRSEC DMR-1119826. This research was supported by the Army Research Office (ARO) under Grant No.W911NF-14-1-0011. J.B. and K.C. acknowledge partial support from the ARO Grant No. W911NF-16-1-0349. P.R. acknowledges partial support from the Air Force Office Scientific Research under Grant No. FA9550-15-1-0015. S. S. acknowledges partial support from the ARO Grant No. W911NF-14-1-0563. C.A. acknowledges support from the NSF Graduate Research Fellowship under Grant No. DGE-1122492.
\end{acknowledgments}

\bibliography{bib_QZD}

\newpage

\clearpage

\section{Supplementary Methods}

\renewcommand{\thefigure}{S\arabic{figure}}

\setcounter{figure}{0}

\subsection{Experimental setup}

\begin{figure*}
\includegraphics{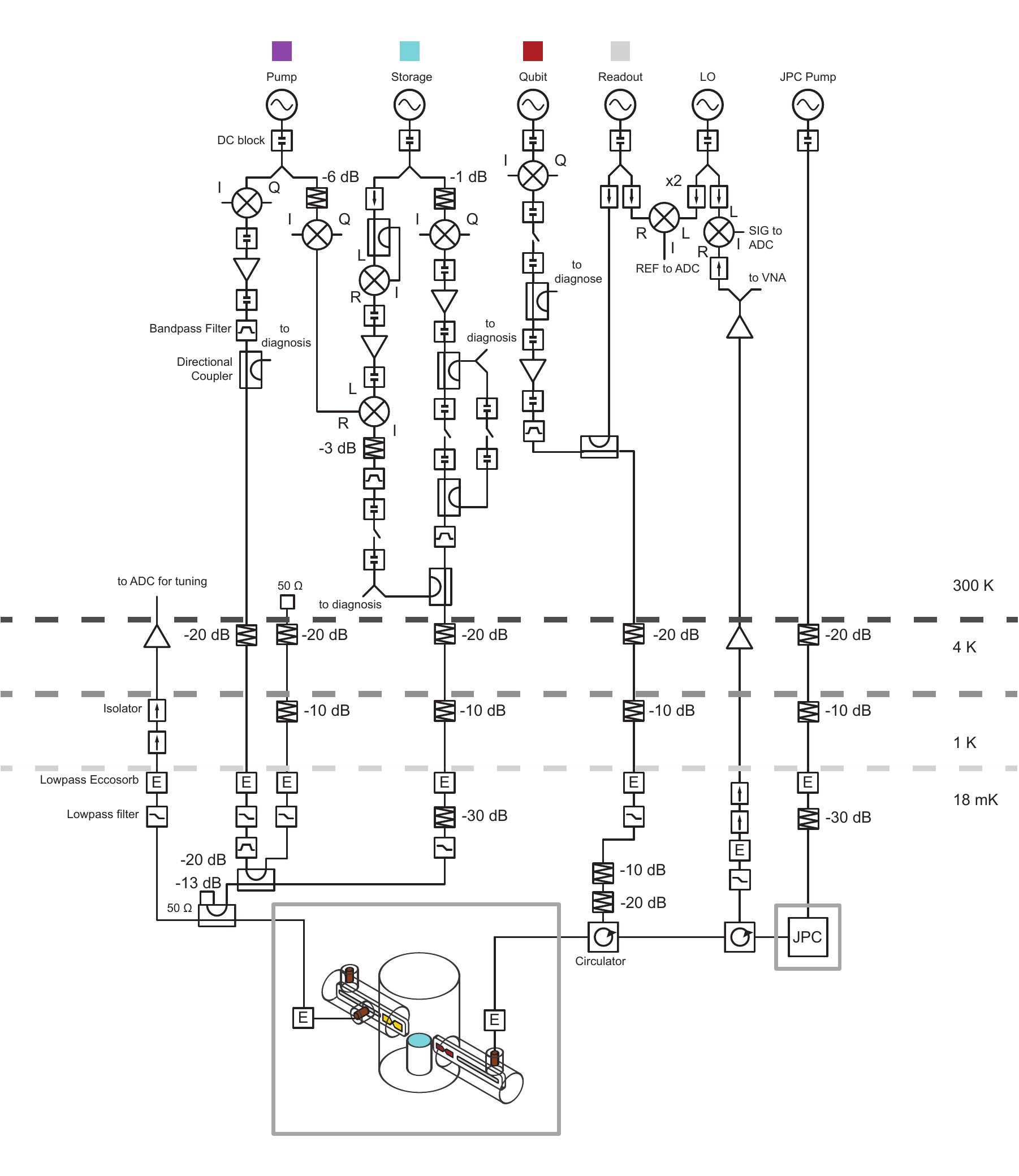}
\caption{Experimental setup. See text of first section. }
\label{fig:OpticalTable}
\end{figure*}

The detailed setup is shown on Fig. S\ref{fig:OpticalTable}. The top of this figure (above the 300K dashed line) shows the relevant room temperature electronics while the bottom half shows the wiring of the dilution refrigerator. The measurement setup is on the right hand side of the figure. Two generators (Readout and local oscillator (LO)) were used to perform a heterodyne measurement on the Wigner transmon. We performed single shot measurements of the state of the transmon qubit using a Josephson Parametric Converter as a parametric amplifier. The two middle branches were used to control the two quadratures of the signal addressing the transmon qubit and the storage (crimson and cyan). The storage branch served 3 different purposes. First, it was used to perform fast displacements on the storage cavity (both for state preparation and to measure the Wigner function). Second we used it for the slow drive that performs the gate. To this end, we used directional couplers to create an additional, strongly attenuated, path on the right side of the storage branch. Finally the left side of the storage branch was part of an interferometer that created the reservoir drive by mixing the frequency-doubled storage tone with the pump tone. This way, the reservoir drive was phase-locked to the pump tone and to the storage tone, such that drifts in the phase of a generator were not affecting the experiment. We also had control over the quadratures of the pump tone and the reservoir drive. Those 4 controls made it possible to sweep the phase and the frequency of the relevant modes of the experiment and thus simplified the way we tuned it (as shown in next sections). Inside the dilution refrigerator, on the side of the pump, we used a combination of non-dissipative elements that combined drives of very different powers and attenuated the pump without warming up the base plate. We used two directional couplers that combined the storage and reservoir drives with the pump tone and attenuated the pump tone by sending 99\% of its power back to the 4K stage of the dilution refrigerator on another line. This setup also provided a way to diagnose the side of the reservoir by measuring in reflection.

\subsection{Hamiltonian and parameters}

The Hamiltonian of our system, when it is not driven, is well described by the usual circuit QED Hamiltonian \cite{Girvin2012} containing Kerr effect $\chi_{ii}/2(\ba_i^{\dag 2}\ba_i^2$) and cross-Kerr $\chi_{ij}(\ba_i^\dag\ba_i)(\ba_j^\dag\ba_j)$. The coupling between the reservoir and the transmon qubit has been designed to be small. Our measurements did not reveal a coupling that would be bigger than the linewidth of either mode. We thus neglected any coupling between those two modes in our model. The coupling between the storage and the transmon qubit was large enough (\SI{2.29}{\mega \hertz}) to perform a fast parity measurement \cite{Davidovich1997}\cite{Haroche2002} (\SI{218}{\nano \second} $\ll T_1, T_2$). The coupling between the reservoir and the storage was such that we reached a large value for $\kappa_2$ compared to $\kappa_1$ while having a Purcell limit on the lifetime of the storage that was high enough (here \SI{92}{\micro \second}) to see oscillations in the parity before the decoherence of cat states. In Table S\ref{table:freq} we give the coherence times of each mode. The lifetime of an electromagnetic mode stored in an aluminium post cavity is usually of order \SI{1}{\milli \second} \cite{Schoelkopf2013_2}. We attribute our lower lifetime to the Purcell effect from the reservoir to the storage cavity. We attribute the finite dephasing time of the storage cavity to the finite temperature of the modes it was coupled to, particularly the reservoir (see next section). Table S\ref{table:param} shows the measured system parameters. The only parameter not accessible via standard techniques (see supplement of \cite{Devoret2015}) was the coupling between the storage and the reservoir. As we were not in the photon number splitting regime, we could not deduce this quantity using a spectroscopy experiment. Instead, we used a known quantity, the anharmonicity of the reservoir, in order to measure it. When we applied the pump tone to the reservoir, the frequencies of both the reservoir and the storage underwent a Stark-shift $\Delta_R$ and $\Delta_S$ respectively. From the full system Hamiltonian we derived $\chi_{RS} = 2\chi_{RR}\Delta_S/\Delta_R$ \cite{Devoret2015}. As $\chi_{RR}$ was known by inducing two-photon transition between $\ket{g}_R$ to $\ket{f}_R$, we deduced $\chi_{RS}$ (see Table S\ref{table:param}).

\subsection{Phase-flips characterization}

In our encoding, phase-flips correspond to leakage between the coherent states $\ket{\pm{\alpha_\infty}}$. They were measured by looking at the difference of the Wigner functions at points $\pm {\alpha_\infty}$ after initializing the storage cavity in a coherent state $\ket{{\alpha_\infty}}$.  On Fig. S\ref{fig:phase_flips} we show that if the temperature of our system had been 0K, our manifold stabilization would have protected the encoding against phase-flips exponentially with the average number of photons. The actual experiment revealed a much faster phase decay that did not seem to depend on the number of photons of the cat state. We reproduced this decay by introducing a finite thermal population in the reservoir mode in our simulations (see section on simulations). We used this dependence as a way to evaluate the thermal population of the reservoir mode for simulations of our experiment such as on Fig. 3 of the main paper. 

\subsection{Tuning the frequency matching condition}

The experiment required a precise frequency matching condition for the frequency of the pump tone: $f_P = 2f_S - f_R$. The Hamiltonian of 2 modes coupled through a Josephson junction predicts that when a pump tone is applied, the frequency of each mode is renormalized due to a Stark-shift, which complicated the tuning phase of our experiment. We proceeded by fixing the frequency of the pump tone a little bit above the bare frequency matching condition (to account for the Stark-shift) and then swept its amplitude. We see on Fig. S\ref{fig:Cavity_Stark_shift} that the frequency of the storage cavity moved (and so did the frequency of the reservoir) until the frequency matching condition was met. This condition was obtained when we observed an anti-crossing in the storage spectroscopy. On Fig. S\ref{fig:drivefreq_vs_pumpamp} we looked at the overlap of the state of the storage with Fock state $\ket{0}$ while we were sweeping the amplitude of the pump tone (x-axis) and the frequency of the reservoir drive (y-axis), without sending any signal at the frequency of the storage. When the two-photon conversion happened, the average photon number in the storage was not 0. We see that at the same amplitude as the previous anti-crossing there was an efficient conversion of reservoir photons into pairs of storage photons over a wide range of frequencies for the reservoir drive. 

\subsection{Simulations}

In Fig. 3 of the main paper, we used Python and the open source library QuTip \cite{qutip} to simulate the time evolution of our system. The Hilbert space that we considered was composed of the reservoir (as a three-level system) and the storage (harmonic oscillator with a truncation at 30 levels). We decided to neglect the effect of the transmon qubit for two reasons. First the experiment was performed after checking that the transmon qubit was in its ground state. Second, repeated measurements of the transmon qubit in its equilibrium state showed a thermal population of only 5\% while its $T_1$ was \SI{70}{\micro \second}. The timescale corresponding to jumps towards the first excited state thus was \cite{Haroche2006} $T_1/\bar{n}_{th} =$ \SI{1.4}{\milli \second} which was enough to consider that the transmon qubit would stay in its ground state. We simulated the following Hamiltonian:

\begin{align*}
\frac{\bH_{RS}}{\hbar} &= \left(g\ba_S^2\ba_R^\dag + g^*\ba_S^{\dag 2}\ba_R\right) + \left(\epsilon_R\ba_R^\dag + \epsilon_R^*\ba_R\right) + \left(\epsilon\ba_S^\dag + \epsilon^*\ba_S\right)\\
       &\quad - \frac{\chi_{RR}}{2}\ba_R^{\dag 2}\ba_R^2 - \frac{\chi_{SS}}{2}\ba_S^{\dag 2}\ba_S^2 - \chi_{RS}(\ba_R^\dag\ba_R)(\ba_S^\dag\ba_S) \\
       &\quad + \Delta_R \ba_R^\dag\ba_R + \frac{\Delta_R + \Delta_P}{2} \ba_S^\dag\ba_S. 
\end{align*}

\begin{figure*}
\includegraphics{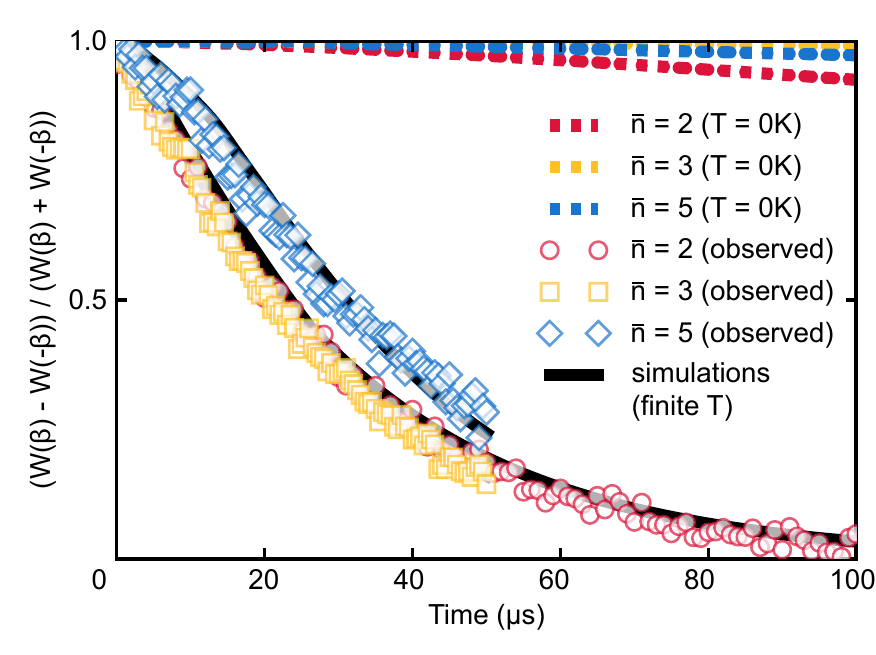}
\caption{Temperature dependent phase-flip error. Normalized leakage from the initial coherent state $\ket{{\alpha_\infty}}$ to the state of opposite phase as a function of time. The dashed red, yellow and blue curves give the theoretical behavior at zero temperature for $\bar{n}=$ 2, 3 and 5 respectively. The corresponding markers show the experimentally observed values. Each of the latter curves is well reproduced by simulation (solid black lines) for a thermal population of 1.5\% for $\bar{n}=$ 2 and 3 and 2\% for $\bar{n}=$ 5 in the reservoir mode. The data and the corresponding simulations are taken at a fixed frequency for the microwave pumps. As the frequency matching condition is not optimized for each $\bar{n}$, the theory curve for $\bar{n}$ = 3 has a lower leakage rate than $\bar{n}$ = 5.}
\label{fig:phase_flips}
\end{figure*}

\begin{figure*}
\includegraphics{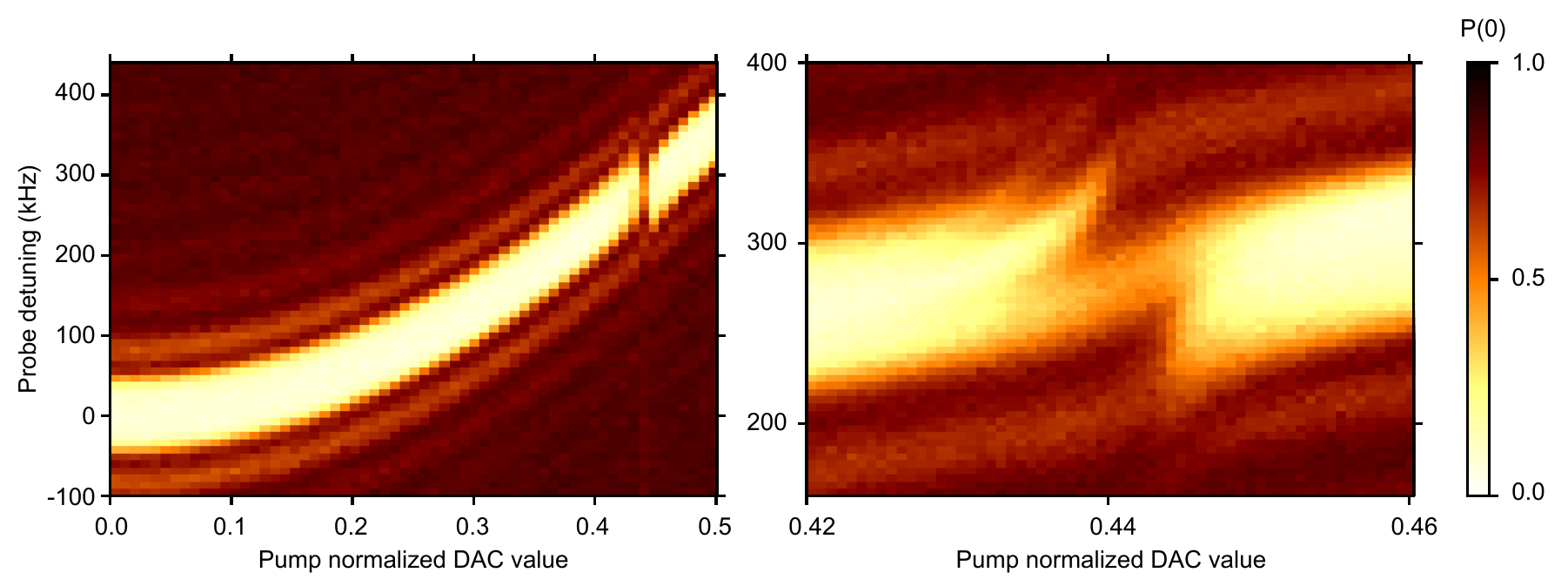}
\caption{Cavity Stark shift. We do a spectroscopy of the storage cavity while applying the pump with different strengths. We send a \SI{17}{\micro \second} square pulse and measure the overlap of the resulting storage state with the Fock state $\ket{0}$. The frequency of the storage pulse is modulated around the value corresponding to the bare frequency of the storage cavity (given by zero probe detuning). The x-axis shows the amplitude of the pump. The zero means that no tone is sent while 1 corresponds to the maximum amplitude that our control electronics can deliver. The right panel is a zoom of the anti-crossing corresponding to the frequency matching condition being met.}
\label{fig:Cavity_Stark_shift}
\end{figure*}

\begin{figure*}
\includegraphics{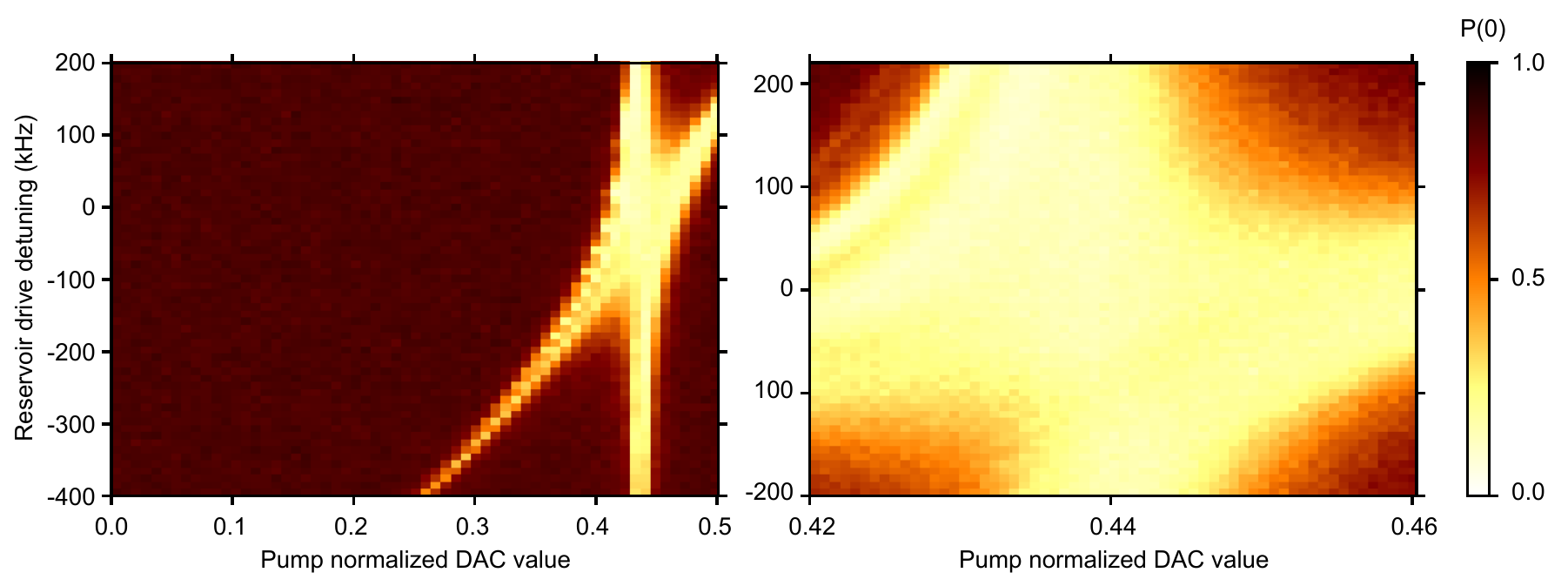}
\caption{Frequency of the reservoir drive versus amplitude of the pump tone. We play the pump tone and the reservoir drive during \SI{100}{\micro \second} and plot the overlap of the resulting storage state with $\ket{0}$ for different frequencies of the reservoir drive and amplitudes for the pump tone. The reservoir drive frequency is expressed by its detuning from the Stark-shifted frequency of the reservoir at the frequency matching condition. The right panel is a zoom on the anti-crossing taking place at that particular point.}
\label{fig:drivefreq_vs_pumpamp}
\end{figure*}

As described in \cite{Devoret2015}, this Hamiltonian is reducible to the one detailed in the main paper. The value of $g$ and of $\epsilon_R$ were deduced from measuring the frequency Stark-shift of the reservoir for a given pump strength and from the size $\bar{n}$ of the stabilized cat states \cite{Devoret2015}. The $\chi_{ij}$ were all measured from independent experiments and their values are given in Table S\ref{table:param}. $\Delta_R$ and $\Delta_P$ were accounting for possible detunings of the microwave pumps from the perfect frequency matching condition. The simulations for Fig. 3 were done at the frequency matching condition $\Delta_R = \Delta_P = 0$. The value of $\epsilon$ was measured from the frequency of the parity oscillations on Fig. 3. The full Lindblad equation is then

\begin{align*}
\frac{d\rho_{RS}}{dt} &= -\frac{i}{\hbar}\left[\bH_{RS}, \rho_{RS}\right] + \frac{\kappa_1}{2}\mathcal{D}\left[\ba_S\right]\rho_{RS} + \frac{1 + \bar{n}_{th}}{2T_{1R}}\mathcal{D}\left[\ba_R\right]\rho_{RS} + \frac{\bar{n}_{th}}{2T_{1R}}\mathcal{D}\left[\ba_R^\dag\right]\rho_{RS},
\end{align*}
where the dissipation timescales are given in Table S\ref{table:freq} and $\bar{n}_{th}$ corresponds to the thermal population of the reservoir mode. We chose a tensor product between a cat state in the storage cavity and a thermal state in the reservoir as initial state. The thermal population of the reservoir was deduced from the phase-flip measurement on Fig. S\ref{fig:phase_flips}. The numerical integration of the differential equation gave a final density matrix for the entire system. In order to compare the result to the data, we took a partial trace over the reservoir mode and then calculated the Wigner function for zero displacement in phase space. We scaled it by the amplitude of our parity measurement for a given cat state and superimposed the resulting curve to the data (see Fig. 3).

%
%\begin{figure*}
%\includegraphics{parity_vs_angle.jpg}
%\caption{\textbf{Tuning the phase of the drive.} }
%\label{fig:parity_vs_angle}
%\end{figure*}

\clearpage

\section{Supplementary Tables}

\begin{table}[h]
\centering
\caption{Frequencies and coherence times of the experimental design.}
\medskip
\begin{tabular}{cccc}
\hline
Mode & Frequency (\SI{}{\giga \hertz}) & $T_1$ (\SI{}{\micro \second}) & $T_2$ (\SI{}{\micro \second})\\
\hline
Qubit & 5.89064 & 70 & 30\\
Storage & 8.10451 & 92 & 40\\
Reservoir & 6.6373 & 0.317 & -\\
\hline
\label{table:freq}
\end{tabular}
\end{table}

\begin{table}[h]
\centering
\caption{Parameters of the Hamiltonian of the experimental device.}
\medskip
\begin{tabular}{c|cccc}
\hline
$\chi_{ij} / 2 \pi$ (\SI{}{\mega \hertz}) & Qubit & Storage & Reservoir\\
\hline
Qubit & 268 & - & -\\
Storage & 2.29 & $\sim$ 0.003 & -\\
Reservoir & - & 0.471 & 86\\
\hline
\label{table:param}
\end{tabular}
\end{table}

\clearpage

\end{document}